\documentclass[longauth]{aa} 
\usepackage[varg]{txfonts}
\usepackage{natbib}
\usepackage{url}
\usepackage{hyperref}
\usepackage{xcolor}
\usepackage{ulem}

\usepackage{xspace}

\usepackage{upgreek}
\usepackage{graphicx}   

\newcommand{\be}{\begin{equation}}
\newcommand{\ee}{\end{equation}}
\newcommand{\bea}{\begin{eqnarray}}
\newcommand{\eea}{\end{eqnarray}}

\newcommand {\lum}{erg s$^{-1}$}

\newcommand {\gx}{GX~301$-$2\xspace}

\begin{document}

\title{X-ray polarimetry of the accreting pulsar GX 301$-$2}
 \titlerunning{X-ray polarimetry of GX 301$-$2}

\authorrunning{V.~F. Suleimanov et al.}

\author{Valery~F.~Suleimanov\inst{\ref{in:Tub}}
\and Sofia~V.~Forsblom \inst{\ref{in:UTU}}
\and Sergey~S.~Tsygankov \inst{\ref{in:UTU}}
\and Juri Poutanen \inst{\ref{in:UTU}}
\and Victor~Doroshenko\inst{\ref{in:Tub}}
\and Rosalia~Doroshenko\inst{\ref{in:Tub}}  
\and Fiamma~Capitanio \inst{\ref{in:INAF-IAPS}} 
\and Alessandro~Di~Marco \inst{\ref{in:INAF-IAPS}} 
\and Denis~Gonz{\'a}lez-Caniulef \inst{\ref{in:IRAP}} 
\and Jeremy~Heyl \inst{\ref{in:UBC}}
\and Fabio~La~Monaca \inst{\ref{in:INAF-IAPS}} 
\and Alexander~A.~Lutovinov \inst{\ref{in:IKI}} 
\and Sergey~V.~Molkov\inst{\ref{in:IKI}}
\and Christian Malacaria \inst{\ref{in:ISSI}}
\and Alexander~A.~Mushtukov \inst{\ref{in:Oxford}} 
\and Andrey~E.~Shtykovsky \inst{\ref{in:IKI},\ref{in:HSE}} 
\and Iv\'an~Agudo \inst{\ref{in:CSIC-IAA}}
\and Lucio~A.~Antonelli \inst{\ref{in:INAF-OAR},\ref{in:ASI-SSDC}} 
\and Matteo~Bachetti \inst{\ref{in:INAF-OAC}} 
\and Luca~Baldini  \inst{\ref{in:INFN-PI},      \ref{in:UniPI}} 
\and Wayne~H.~Baumgartner  \inst{\ref{in:NASA-MSFC}}
\and Ronaldo~Bellazzini  \inst{\ref{in:INFN-PI}} 
\and Stefano~Bianchi \inst{\ref{in:UniRoma3}}  
\and Stephen~D.~Bongiorno \inst{\ref{in:NASA-MSFC}} 
\and Raffaella~Bonino  \inst{\ref{in:INFN-TO},\ref{in:UniTO}}
\and Alessandro~Brez  \inst{\ref{in:INFN-PI}} 
\and Niccol\`{o}~Bucciantini 
\inst{\ref{in:INAF-Arcetri},\ref{in:UniFI},\ref{in:INFN-FI}}
\and Simone~Castellano \inst{\ref{in:INFN-PI}} 
\and Elisabetta~Cavazzuti \inst{\ref{in:ASI}} 
\and Chien-Ting~Chen \inst{\ref{in:USRA-MSFC}}
\and Stefano~Ciprini \inst{\ref{in:INFN-Roma2},\ref{in:ASI-SSDC}}
\and Enrico~Costa \inst{\ref{in:INAF-IAPS}} 
\and Alessandra~De~Rosa \inst{\ref{in:INAF-IAPS}} 
\and Ettore~Del~Monte \inst{\ref{in:INAF-IAPS}} 
\and Laura~Di~Gesu \inst{\ref{in:ASI}} 
\and Niccol\`{o}~Di~Lalla \inst{\ref{in:Stanford}}
\and Immacolata~Donnarumma \inst{\ref{in:ASI}}
\and Michal~Dov\v{c}iak \inst{\ref{in:CAS-ASU}}
\and Steven~R.~Ehlert \inst{\ref{in:NASA-MSFC}}  
\and Teruaki~Enoto \inst{\ref{in:RIKEN}}
\and Yuri~Evangelista \inst{\ref{in:INAF-IAPS}}
\and Sergio~Fabiani \inst{\ref{in:INAF-IAPS}}
\and Riccardo~Ferrazzoli \inst{\ref{in:INAF-IAPS}} 
\and Javier~A.~Garcia \inst{\ref{in:Caltech}}
\and Shuichi~Gunji \inst{\ref{in:Yamagata}} 
\and Kiyoshi~Hayashida \inst{\ref{in:Osaka}}\thanks{Deceased.}  
\and Wataru~Iwakiri \inst{\ref{in:Chiba}} 
\and Svetlana~G.~Jorstad \inst{\ref{in:BU},\ref{in:SPBU}} 
\and Philip~Kaaret \inst{\ref{in:NASA-MSFC}}  
\and Vladimir~Karas \inst{\ref{in:CAS-ASU}}
\and Fabian~Kislat \inst{\ref{in:UNH}} 
\and Takao~Kitaguchi  \inst{\ref{in:RIKEN}} 
\and Jeffery~J.~Kolodziejczak \inst{\ref{in:NASA-MSFC}} 
\and Henric~Krawczynski  \inst{\ref{in:WUStL}}
\and Luca~Latronico  \inst{\ref{in:INFN-TO}} 
\and Ioannis~Liodakis \inst{\ref{in:FINCA}}
\and Simone~Maldera \inst{\ref{in:INFN-TO}}  
\and Alberto~Manfreda \inst{\ref{INFN-NA}}
\and Fr\'{e}d\'{e}ric~Marin \inst{\ref{in:Strasbourg}} 
\and Andrea~Marinucci \inst{\ref{in:ASI}} 
\and Alan~P.~Marscher \inst{\ref{in:BU}} 
\and Herman~L.~Marshall \inst{\ref{in:MIT}}
\and Francesco~Massaro \inst{\ref{in:INFN-TO},\ref{in:UniTO}} 
\and Giorgio~Matt  \inst{\ref{in:UniRoma3}}  
\and Ikuyuki~Mitsuishi \inst{\ref{in:Nagoya}} 
\and Tsunefumi~Mizuno \inst{\ref{in:Hiroshima}} 
\and Fabio~Muleri \inst{\ref{in:INAF-IAPS}} 
\and Michela~Negro \inst{\ref{in:UMBC},\ref{in:NASA-GSFC},\ref{in:CRESST}} 
\and Chi-Yung~Ng \inst{\ref{in:HKU}}
\and Stephen~L.~O'Dell \inst{\ref{in:NASA-MSFC}}  
\and Nicola~Omodei \inst{\ref{in:Stanford}}
\and Chiara~Oppedisano \inst{\ref{in:INFN-TO}}  
\and Alessandro~Papitto \inst{\ref{in:INAF-OAR}}
\and George~G.~Pavlov \inst{\ref{in:PSU}}
\and Abel~L.~Peirson \inst{\ref{in:Stanford}}
\and Matteo~Perri \inst{\ref{in:ASI-SSDC},\ref{in:INAF-OAR}}
\and Melissa~Pesce-Rollins \inst{\ref{in:INFN-PI}} 
\and Pierre-Olivier~Petrucci \inst{\ref{in:Grenoble}}
\and Maura~Pilia \inst{\ref{in:INAF-OAC}} 
\and Andrea~Possenti \inst{\ref{in:INAF-OAC}} 
\and Simonetta~Puccetti \inst{\ref{in:ASI-SSDC}}
\and Brian~D.~Ramsey \inst{\ref{in:NASA-MSFC}}  
\and John~Rankin \inst{\ref{in:INAF-IAPS}} 
\and Ajay~Ratheesh \inst{\ref{in:INAF-IAPS}} 
\and Oliver~J.~Roberts \inst{\ref{in:USRA-MSFC}}
\and Roger~W.~Romani \inst{\ref{in:Stanford}}
\and Carmelo~Sgr\`o \inst{\ref{in:INFN-PI}}  
\and Patrick~Slane \inst{\ref{in:CfA}}  
\and Paolo~Soffitta \inst{\ref{in:INAF-IAPS}} 
\and Gloria~Spandre \inst{\ref{in:INFN-PI}} 
\and Douglas~A.~Swartz \inst{\ref{in:USRA-MSFC}}
\and Toru~Tamagawa \inst{\ref{in:RIKEN}}
\and Fabrizio~Tavecchio \inst{\ref{in:INAF-OAB}}
\and Roberto~Taverna \inst{\ref{in:UniPD}}  
\and Yuzuru~Tawara \inst{\ref{in:Nagoya}}
\and Allyn~F.~Tennant \inst{\ref{in:NASA-MSFC}}  
\and Nicholas~E.~Thomas \inst{\ref{in:NASA-MSFC}}  
\and Francesco~Tombesi  \inst{\ref{in:UniRoma2},\ref{in:INFN-Roma2},\ref{in:UMd}}
\and Alessio~Trois \inst{\ref{in:INAF-OAC}}
\and Roberto~Turolla \inst{\ref{in:UniPD},\ref{in:MSSL}}
\and Jacco~Vink \inst{\ref{in:Amsterdam}}
\and Martin~C.~Weisskopf \inst{\ref{in:NASA-MSFC}} 
\and Kinwah~Wu \inst{\ref{in:MSSL}}
\and Fei~Xie \inst{\ref{in:GSU},\ref{in:INAF-IAPS}}
\and Silvia~Zane  \inst{\ref{in:MSSL}}
}

\institute{
Institut f\"ur Astronomie und Astrophysik, Universit\"at T\"ubingen, Sand 1, D-72076 T\"ubingen, Germany \label{in:Tub}
\\ \email{suleimanov@astro.uni-tuebingen.de}
\and
Department of Physics and Astronomy, FI-20014 University of Turku,  Finland \label{in:UTU} 
\and  INAF Istituto di Astrofisica e Planetologia Spaziali, Via del Fosso del Cavaliere 100, 00133 Roma, Italy \label{in:INAF-IAPS}
\and Institut de Recherche en Astrophysique et Plan\'etologie, UPS-OMP, CNRS, CNES, 9 avenue du Colonel Roche, BP 44346 31028, Toulouse CEDEX 4, France \label{in:IRAP}
\and 
University of British Columbia, Vancouver, BC V6T 1Z4, Canada \label{in:UBC}
\and Space Research Institute (IKI) of Russian Academy of Sciences, Prosoyuznaya ul 84/32, 117997 Moscow, Russian Federation \label{in:IKI}
\and International Space Science Institute, Hallerstrasse 6, 3012 Bern, Switzerland \label{in:ISSI}
\and Astrophysics, Department of Physics, University of Oxford, Denys Wilkinson Building, Keble Road, Oxford OX1 3RH, UK \label{in:Oxford}
\and
National Research University Higher School of Economics, Faculty of Physics, 101000, Myasnitskaya ul. 20, Moscow, Russia
\label{in:HSE}
\and 
Instituto de Astrof\'{i}sicade Andaluc\'{i}a -- CSIC, Glorieta de la Astronom\'{i}a s/n, 18008 Granada, Spain \label{in:CSIC-IAA}
\and 
INAF Osservatorio Astronomico di Roma, Via Frascati 33, 00040 Monte Porzio Catone (RM), Italy \label{in:INAF-OAR}  
\and 
Space Science Data Center, Agenzia Spaziale Italiana, Via del Politecnico snc, 00133 Roma, Italy \label{in:ASI-SSDC}
 \and
INAF Osservatorio Astronomico di Cagliari, Via della Scienza 5, 09047 Selargius (CA), Italy  \label{in:INAF-OAC}
\and 
Istituto Nazionale di Fisica Nucleare, Sezione di Pisa, Largo B. Pontecorvo 3, 56127 Pisa, Italy \label{in:INFN-PI}
\and  
Dipartimento di Fisica, Universit\`{a} di Pisa, Largo B. Pontecorvo 3, 56127 Pisa, Italy \label{in:UniPI} 
\and 
NASA Marshall Space Flight Center, Huntsville, AL 35812, USA \label{in:NASA-MSFC}
\and 
Dipartimento di Matematica e Fisica, Universit\`a degli Studi Roma Tre, via della Vasca Navale 84, 00146 Roma, Italy  \label{in:UniRoma3}
\and  
Istituto Nazionale di Fisica Nucleare, Sezione di Torino, Via Pietro Giuria 1, 10125 Torino, Italy  \label{in:INFN-TO}      
\and  
Dipartimento di Fisica, Universit\`{a} degli Studi di Torino, Via Pietro Giuria 1, 10125 Torino, Italy \label{in:UniTO} 
\and   
INAF Osservatorio Astrofisico di Arcetri, Largo Enrico Fermi 5, 50125 Firenze, Italy 
\label{in:INAF-Arcetri} 
\and  
Dipartimento di Fisica e Astronomia, Universit\`{a} degli Studi di Firenze, Via Sansone 1, 50019 Sesto Fiorentino (FI), Italy \label{in:UniFI} 
\and   
Istituto Nazionale di Fisica Nucleare, Sezione di Firenze, Via Sansone 1, 50019 Sesto Fiorentino (FI), Italy \label{in:INFN-FI}
\and 
Agenzia Spaziale Italiana, Via del Politecnico snc, 00133 Roma, Italy \label{in:ASI}
\and 
Science and Technology Institute, Universities Space Research Association, Huntsville, AL 35805, USA \label{in:USRA-MSFC}
\and 
Istituto Nazionale di Fisica Nucleare, Sezione di Roma ``Tor Vergata'', Via della Ricerca Scientifica 1, 00133 Roma, Italy 
 \label{in:INFN-Roma2}
\and 
Department of Physics and Kavli Institute for Particle Astrophysics and Cosmology, Stanford University, Stanford, California 94305, USA  \label{in:Stanford}
\and 
Universit\'{e} Grenoble Alpes, CNRS, IPAG, 38000 Grenoble, France \label{in:Grenoble}
\and 
Astronomical Institute of the Czech Academy of Sciences, Boční II 1401/1, 14100 Praha 4, Czech Republic \label{in:CAS-ASU}
\and 
RIKEN Cluster for Pioneering Research, 2-1 Hirosawa, Wako, Saitama 351-0198, Japan \label{in:RIKEN}
\and 
Cahill Center for Astronomy and Astrophysics, California Institute of Technology, Pasadena, CA 91125, USA \label{in:Caltech}
\and 
Yamagata University,1-4-12 Kojirakawa-machi, Yamagata-shi 990-8560, Japan \label{in:Yamagata}
\and 
Osaka University, 1-1 Yamadaoka, Suita, Osaka 565-0871, Japan \label{in:Osaka}
\and 
International Center for Hadron Astrophysics, Chiba University, Chiba 263-8522, Japan \label{in:Chiba}
\and
Institute for Astrophysical Research, Boston University, 725 Commonwealth Avenue, Boston, MA 02215, USA \label{in:BU} 
\and 
Department of Astrophysics, St. Petersburg State University, Universitetsky pr. 28, Petrodvoretz, 198504 St. Petersburg, Russia \label{in:SPBU} 
\and 
Department of Physics and Astronomy and Space Science Center, University of New Hampshire, Durham, NH 03824, USA \label{in:UNH} 
\and 
Physics Department and McDonnell Center for the Space Sciences, Washington University in St. Louis, St. Louis, MO 63130, USA \label{in:WUStL}
\and 
Finnish Centre for Astronomy with ESO,  20014 University of Turku, Finland \label{in:FINCA}
\and 
Istituto Nazionale di Fisica Nucleare, Sezione di Napoli, Strada Comunale Cinthia, 80126 Napoli, Italy \label{INFN-NA}
\and 
Universit\'{e} de Strasbourg, CNRS, Observatoire Astronomique de Strasbourg, UMR 7550, 67000 Strasbourg, France \label{in:Strasbourg}
\and 
MIT Kavli Institute for Astrophysics and Space Research, Massachusetts Institute of Technology, 77 Massachusetts Avenue, Cambridge, MA 02139, USA \label{in:MIT}
\and 
Graduate School of Science, Division of Particle and Astrophysical Science, Nagoya University, Furo-cho, Chikusa-ku, Nagoya, Aichi 464-8602, Japan \label{in:Nagoya}
\and 
Hiroshima Astrophysical Science Center, Hiroshima University, 1-3-1 Kagamiyama, Higashi-Hiroshima, Hiroshima 739-8526, Japan \label{in:Hiroshima}
\and  
University of Maryland, Baltimore County, Baltimore, MD 21250, USA \label{in:UMBC}
\and 
NASA Goddard Space Flight Center, Greenbelt, MD 20771, USA  \label{in:NASA-GSFC}
\and 
Center for Research and Exploration in Space Science and Technology, NASA/GSFC, Greenbelt, MD 20771, USA  \label{in:CRESST}
\and 
Department of Physics, University of Hong Kong, Pokfulam, Hong Kong \label{in:HKU}
\and 
Department of Astronomy and Astrophysics, Pennsylvania State University, University Park, PA 16801, USA \label{in:PSU}
\and 
Center for Astrophysics, Harvard \& Smithsonian, 60 Garden St, Cambridge, MA 02138, USA \label{in:CfA} 
\and 
INAF Osservatorio Astronomico di Brera, via E. Bianchi 46, 23807 Merate (LC), Italy \label{in:INAF-OAB}
\and 
Dipartimento di Fisica e Astronomia, Universit\`{a} degli Studi di Padova, Via Marzolo 8, 35131 Padova, Italy \label{in:UniPD}
\and
Dipartimento di Fisica, Universit\`{a} degli Studi di Roma ``Tor Vergata'', Via della Ricerca Scientifica 1, 00133 Roma, Italy \label{in:UniRoma2}
\and
Department of Astronomy, University of Maryland, College Park, Maryland 20742, USA \label{in:UMd}
\and 
Mullard Space Science Laboratory, University College London, Holmbury St Mary, Dorking, Surrey RH5 6NT, UK \label{in:MSSL}
\and 
Anton Pannekoek Institute for Astronomy \& GRAPPA, University of Amsterdam, Science Park 904, 1098 XH Amsterdam, The Netherlands  \label{in:Amsterdam}
\and 
Guangxi Key Laboratory for Relativistic Astrophysics, School of Physical Science and Technology, Guangxi University, Nanning 530004, China \label{in:GSU}
}


\abstract
{The phase- and energy-resolved polarization measurements of accreting X-ray pulsars (XRPs) allow us to test different theoretical models of their emission, and they also provide an avenue to determine the emission region geometry.
We present the results of the observations of the XRP \gx performed with the \textit{Imaging X-ray Polarimetry Explorer} (\textit{IXPE}).
A persistent XRP, \gx has one of the longest spin periods known: $\sim$680~s. 
A massive hyper-giant companion star Wray 977 supplies mass to the neutron star via powerful stellar winds. 
We did not detect significant polarization in the phase-averaged data when using spectro-polarimetric analysis, with the upper limit on the polarization degree (PD) of 2.3\% (99\% confidence level). 
Using the phase-resolved spectro-polarimetric analysis, we obtained a significant detection of polarization (above 99\% confidence level) in two out of nine phase bins and a marginal detection in three bins, with a PD ranging between $\sim$3\% and $\sim$10\% and a polarization angle varying in a very wide range from $\sim$0\degr\ to $\sim$160\degr.
Using the rotating vector model, we obtained constraints on the pulsar geometry using both phase-binned and unbinned analyses, finding excellent agreement.  
Finally, we discuss possible reasons for a low observed polarization in \gx. 
}


\maketitle



\section{Introduction}

Accreting X-ray pulsars \citep[XRPs; see][for a recent review]{Mushtukov22} are highly magnetized neutron stars (NSs) with a surface magnetic field strength of $10^{12}$--$10^{13}$\,G orbiting early-type (typically O-B) donor stars. 
A pulsar accretes matter lost by the donor star through the stellar wind or, in the case of Be systems, a decretion disk.
In some systems, an accretion disk around the NS may also form. 
The accretion flow in the vicinity of the NS surface is governed by the magnetic field, and the accreting matter forms hotspots near the NS magnetic poles radiating in the X-ray band. 
There are no commonly accepted models of the radiating regions; however, it is clear that at high enough mass accretion rates, the emitting hotspots transform to radially extended accretion columns \citep{1976MNRAS.175..395B, 2012A&A...544A.123B,
2015MNRAS.447.1847M}.   

Radiation generated by plasma in a strong magnetic field was previously expected to be highly polarized at the photon energies below the cyclotron energy  because of the birefringence of magnetized plasma \citep[see, e.g.,][]{2006RPPh...69.2631H,CH1_21}.  
Testing this prediction recently became possible thanks to the \textit{Imaging X-ray Polarimetry Explorer} (\textit{IXPE}), as it allows one to measure X-ray polarization in the 2--8 keV energy band. 
In the first year of its operation, \textit{IXPE} has carried out polarization measurements of several XRPs (\mbox{Her X-1}, \citealt{Doroshenko22}; \mbox{Cen X-3},  \citealt{Tsygankov22}; \mbox{4U 1626$-$67},  \citealt{Marshall22}; \mbox{Vela X-1}, \citealt{Forsblom2023}; \mbox{GRO J1008$-$57}, \citealt{Tsygankov2023}; \mbox{EXO\,2030+375}, \citealt{Malacaria23}; \mbox{X Persei},  \citealt{Mushtukov23}; \mbox{LS\,V\,+44\,17}, \citealt{Doroshenko23}). 
In most cases, the observations revealed a rather low pulse-phase averaged polarization degree (PD).
The reason for the low PD is not completely clear.
A toy model considering the upper layer of an NS atmosphere overheated by accretion has been proposed \citep{Doroshenko22}. 
According to this model, the escaping radiation can be depolarized due to partial mode conversion when passing through the vacuum resonance region situated at the sharp temperature gradient between the cool inner atmospheric layers and the upper overheated layers.
Additionally, variations of the polarization angle (PA) with the pulsar phase can lead to strong depolarization in the total signal.

\begin{figure}
\centering
\includegraphics[width=0.95\columnwidth]{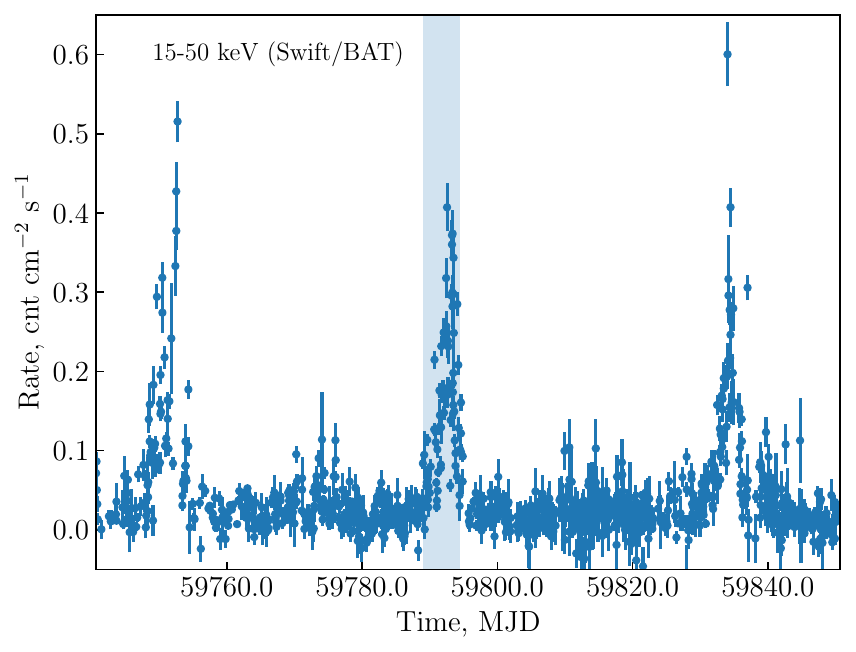}
\caption{Long-term light curve of \gx in the 15--50 keV energy band obtained with the {\it Swift}/BAT monitor. The shaded region represents the time of the \textit{IXPE} observation.  
}
\label{fig:lc}
\end{figure}

The accreting XRP \gx (also associated with 4U 1223$-$62) is one of \textit{IXPE}'s targets.
It is a high-mass X-ray binary system containing a slowly rotating \citep[$P_{\rm spin}\sim$\,680 s,][]{spin76} NS orbiting the hyper-giant donor star Wray 977 \citep{wray76}.
The donor star is very massive (39--53 $M_\odot$) and large ($\sim$\,62 $R_\odot$) with a huge mass-loss rate of $\dot M \sim 10^{-5} M_\odot$\,yr$^{-1}$ via a dense and slow  ($\sim$\,300--400 km\,s$^{-1}$) wind \citep{kaper06}.
The system inclination has been constrained to be in the range of 52\degr--78\degr, with an indication of a preference for an inclination closer to the lower end of the range \citep{2008-Leahy-Kostka}.
A distance to the system of 3.53$^{+0.40}_{-0.52}$\,kpc was measured by \textit{Gaia} \citep{distT18, BJ18}. 
The orbital period of the system is known with a high accuracy: $P_{\rm orb}$=41.482$\pm$ 0.001\,d \citep{Doroshenko.etal.2010}.
The NS orbit is elliptic with an eccentricity $e\sim$\,0.46  \citep{Koh97}, and the X-ray luminosity of the pulsar increases near the periastron, reaching $\sim 10^{37}$\lum (see Fig.~\ref{fig:lc} for the long-term light curve). 
The wind accretion picture in this system is very complicated. 
The outburst maximum is about 1.4\,d before the periastron, and there is a secondary X-ray flux increase near the apoastron at the orbital phase 0.5 \citep{Pravdo95, Koh97}. It was even hypothesized that the NS in the system exhibits retrograde rotation \citep{juhani20}. 

The X-ray spectrum of the system has been investigated by many observatories, including \textit{NuSTAR} \citep[see][and references therein]{Armin19}.
The continuum X-ray spectrum is typical for an accreting XRP and is well described by a power-law model with a Fermi-Dirac cutoff \citep{Furst18} and some additional components.
These include two cyclotron resonant scattering features near 35 and 50 keV  \citep{Furst18}, whose energies are positively correlated with the source X-ray luminosity, including variations over the pulse period \citep{Furst18,Armin19}. 
It is necessary to mention a strong local absorption that is also correlated with the observed luminosity. The absorber column density can be as high as $(5-20)\times 10^{23}$\,cm$^{-2}$
\citep[see, e.g.,][]{Furst18,Armin19, Doroshenko.etal.2010, Ji21}.
Important components of the softer part of the X-ray spectrum are the iron K$\alpha$ and K$\beta$ lines, arising due to a reprocessing of X-rays in the matter surrounding the NS
\citep[see detailed analysis in][]{Ji21}. Their strength is also strongly correlated
with the source luminosity.

The first attempt to measure the X-ray polarization of \gx was performed with balloon-borne hard X-ray polarimeter X-Calibur in the 15--35 keV energy band. This attempt resulted in a non-detection \citep{2020ApJ...891...70A} due to a rather short flight duration. 

In this work, we present the results of \textit{IXPE} observations of \gx. 
The \textit{IXPE} observations were accompanied by simultaneous observations performed with the \textit{Spectrum Roentgen Gamma} (\textit{SRG}) observatory. 

\section{Observations and data reduction}
\label{sec:Obs}

The \textit{IXPE} observatory, a NASA mission in partnership with the Italian space agency \citep[see a detailed description in][]{Weisskopf2022}, was launched by a Falcon 9 rocket on 2021 December 9. 
There are three grazing incidence telescopes onboard the observatory. 
Each telescope is composed of an X-ray mirror  assembly and a polarization-sensitive detector unit (DU) equipped with a gas-pixel detector 
\citep{2021AJ....162..208S,2021APh...13302628B}.
These instruments provide imaging polarimetry in the 2--8 keV energy band with a time resolution better than 10\,$\mu$s.

\begin{figure}
\centering
\includegraphics[width=1.02\columnwidth]{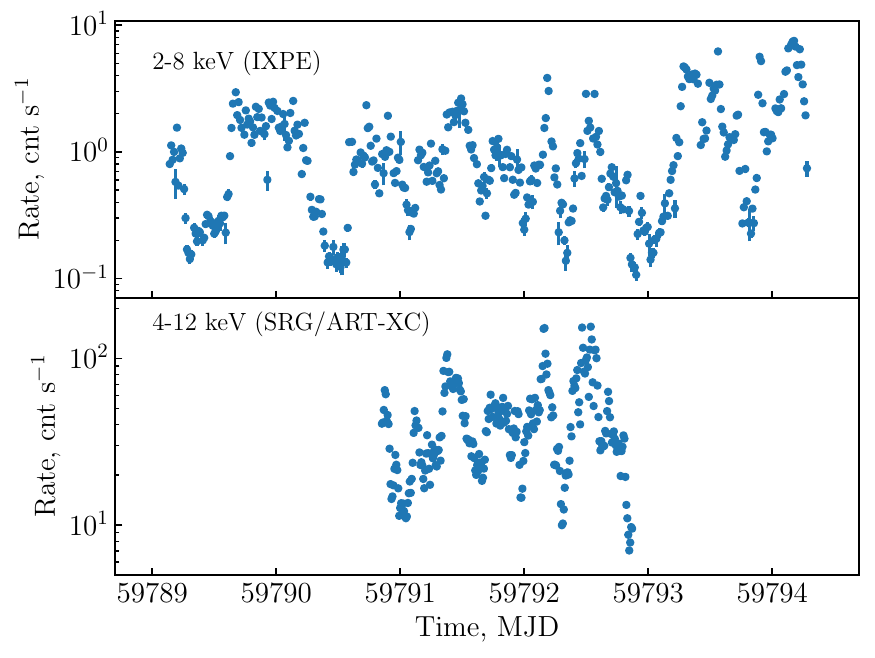}
\caption{Background subtracted light curve of the source in the 2--8 keV energy band observed with \textit{IXPE} (top) and in the 4--12 keV band observed with \textit{SRG}/{ART-XC} (bottom). Data from the three \textit{IXPE} telescopes and the seven {ART-XC} telescopes were combined.
}
\label{fig:lc1}
\end{figure}

Between 2022 July 29 and August 3, \textit{IXPE} observed \gx with a total exposure of $\simeq$290~ks per telescope. 
The data were processed with the {\sc ixpeobssim} package version 30.2.1 \citep{Baldini22} using the CalDB released on 2022 November 17. Before the scientific analysis, a position offset correction and energy calibration were applied.
Source photons were collected in a circular region with a radius $R_{\rm src}=70\arcsec$ centered at the \gx position. 
In the 2--8 keV band, the background makes up about 1.2\% of the total count rate in the source region. 
Because of the low background level and the high source count rate, the background was not subtracted as recommended by \citet{Di_Marco_2023}.
The event arrival times were corrected to the Solar system barycenter using the standard {\tt barycorr} tool from the {\sc ftools} package  and accounting for the effects of binary motion using the orbital parameters from  \citet{Doroshenko.etal.2010}.

\begin{figure}
\centering
\includegraphics[width=0.9\columnwidth, clip]{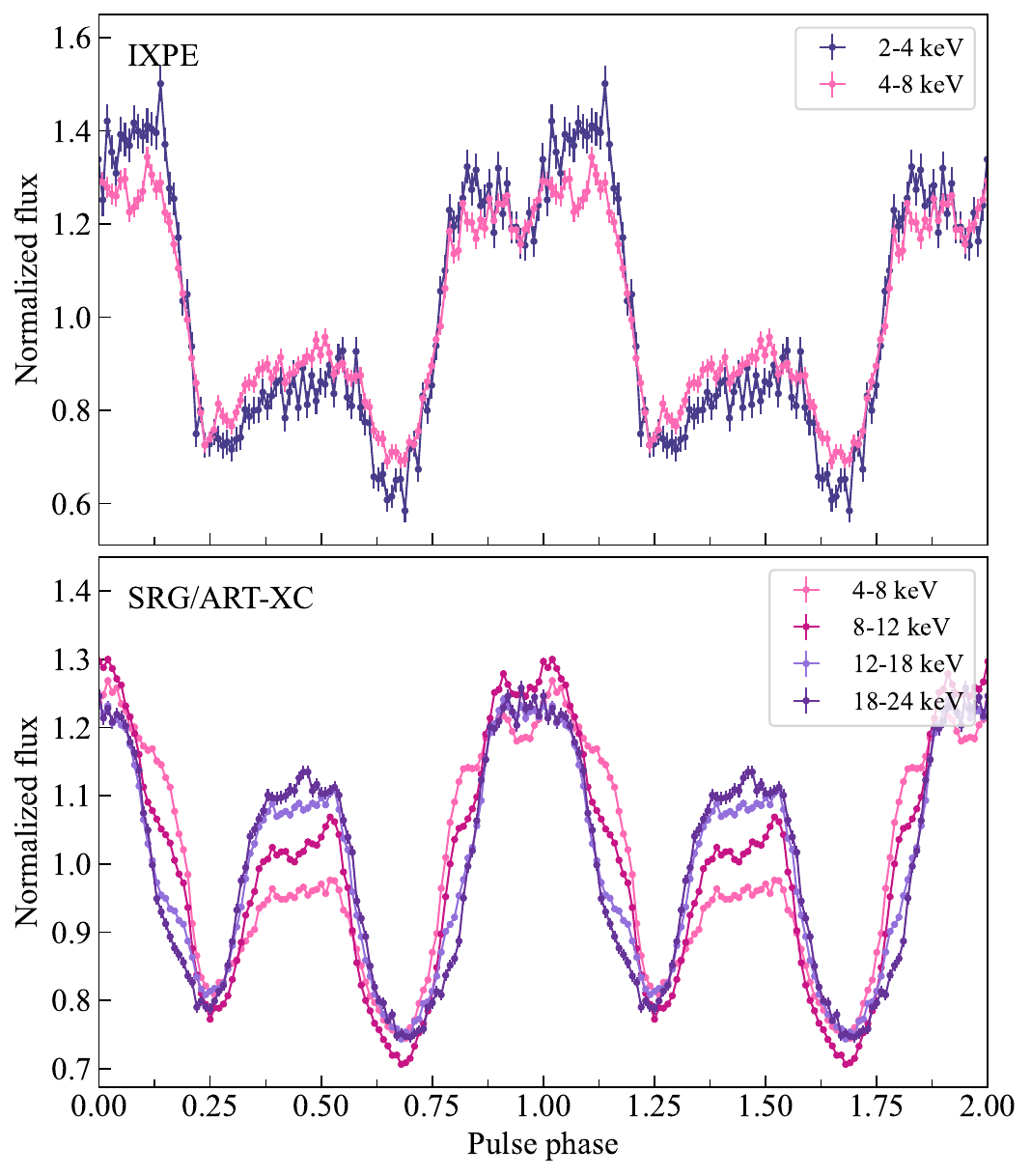}
\caption{Pulse profiles of \gx obtained by \textit{IXPE} and {ART-XC} in different energy bands.}
\label{fig:pulse-profiles-comb}
\end{figure}

The polarimetric parameters of \gx were extracted utilizing the \texttt{pcube} algorithm (\texttt{xpbin} tool) in the \textsc{ixpeobssim} package, which has been implemented following the formalism by \citet{2015-Kislat}. 
For the spectro-polarimetric analysis, the Stokes spectra $I$, $Q$, and $U$ for the source region were prepared using the \texttt{xpbin} tool's \texttt{PHA1}, \texttt{PHA1Q}, and \texttt{PHA1U} algorithms in \textsc{ixpeobssim}, resulting in a data set of nine spectra (three for each DU).

The flux (Stokes parameter $I$) energy spectra were binned to have at least 30 counts per energy channel. 
The same energy binning was also applied to the spectra of the Stokes parameters $Q$ and $U$. 
We applied the unweighted analysis (i.e., taking all events into account independently of the quality of the track reconstruction) of the \textit{IXPE} data. 
All the spectra were fitted with the \textsc{xspec} package \citep{Arn96} (version 12.12.1), which is a part of the standard high-energy astrophysics software suite \texttt{HEASOFT}, using the instrument response functions of version 12 and a $\chi^2$ statistic. 
The uncertainties are given at the 68.3\% confidence level unless stated otherwise.

\begin{figure}
\centering
\includegraphics[width=0.8\columnwidth]{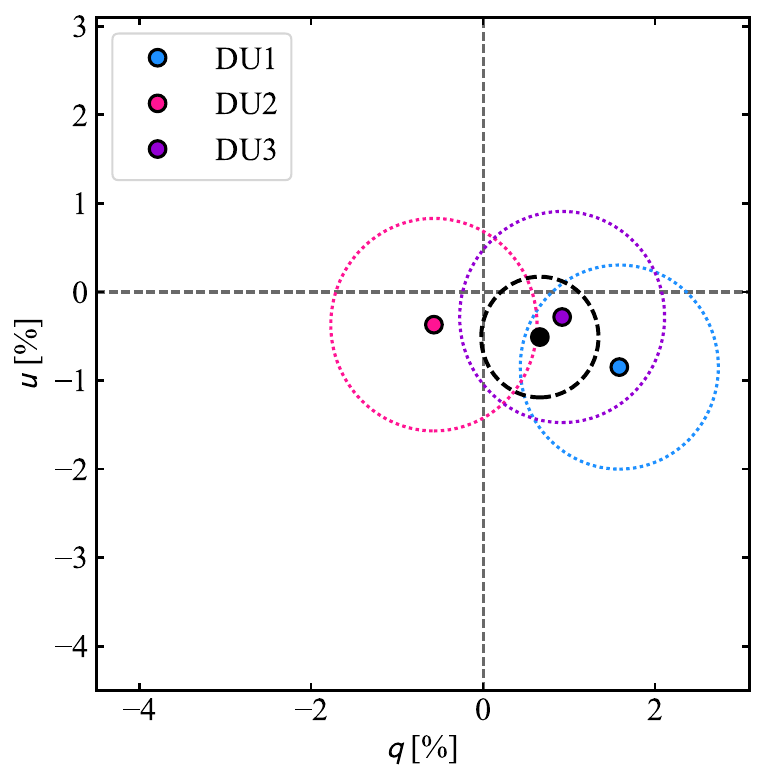}
\caption{Phase-averaged normalized Stokes parameters $q$ and $u$ for each DU and combining the DUs (in black) for the full 2--8 keV energy band. 
The circle size corresponds to the uncertainty at 68\% confidence level.
}
\label{fig:Stokes_averaged}
\end{figure}

The Mikhail Pavlinsky {ART-XC} telescope \citep{Pavlinsky21} on board the \textit{SRG} observatory \citep{Sunyaev21} carried out simultaneous observations of \gx  with \textit{IXPE} starting on 2022 July 30 at  20:25 and ending on August 1 at 20:50 (UTC) with a total exposure of $\sim$160\,ks. 
The {ART-XC} is a grazing incidence focusing X-ray telescope  that provides imaging, timing, and spectroscopy in the 4--30 keV energy range. 
The telescope consists of seven identical modules and has a total effective area of $\sim$450\,cm$^2$ at 8 keV, an angular resolution of 45\arcsec\,, an energy resolution of 1.4 keV at 6 keV, and a timing resolution of 23\,$\mu$s.   
The {ART-XC} data were processed with the analysis software \textsc{artproducts} v1.0 with the CALDB version 20220908.

\begin{table} 
\centering
\caption{Normalized $q$ and $u$ Stokes parameters and PD and PA in different phase bins  obtained from the \texttt{pcube} algorithm.}
\begin{tabular}{ccccc}
    \hline\hline
    Phase & $q$ & $u$ & PD & PA \\
           & (\%) & (\%) & (\%) & (deg) \\
    \hline
    0.000--0.058 & $-10.6\pm3.2$ & $-0.5\pm3.2$& $10.6\pm3.2$ & $-89\pm9$ \\
    0.058--0.193 & $0.1\pm2.0$ & $1.0\pm2.0$   & $1.0^{+2.0}_{-1.0}$& $41\pm58$ \\
    0.193--0.328 & $-0.9\pm2.0$ & $0.9\pm2.0$  & $1.3^{+2.0}_{-1.3}$ & $67\pm44$ \\
    0.328--0.463 & $4.1\pm2.1$ & $2.5\pm2.1$   & $4.8\pm2.1$ & $16\pm13$ \\
    0.463--0.603 & $1.5\pm1.8$ & $0.0\pm1.8$   & $1.5^{+1.8}_{-1.5}$ & $0\pm34$ \\
    0.603--0.733 & $3.8\pm1.7$ & $-0.9\pm1.7$  & $3.9\pm1.7$ & $-7\pm13$ \\
    0.733--0.813 & $0.9\pm2.2$ & $-8.0\pm2.2$  & $8.0\pm2.2$ & $-42\pm8$ \\
    0.813--0.893 & $0.6\pm2.1$ & $-1.2\pm2.1$  & $1.4^{+2.1}_{-1.4}$ & $-32\pm45$ \\
    0.893--1.000 & $-1.5\pm2.0$ & $0.9\pm2.0$  & $1.8^{+2.0}_{-1.8}$ & $75\pm33$ \\  
    \hline
    0.000--1.000 & $0.7\pm0.7$ & $-0.5\pm0.7$  & $0.8\pm0.7$ & $-19\pm23$ \\  
    \hline
\end{tabular}
\tablefoot{The uncertainties are given at the 68.3\% (1$\sigma$) confidence level. }
\label{table:phase-res-PCUBE}
\end{table}

\section{Results}
\label{sec:Res}

\subsection{Light curve and pulse profile}

Between 2022 July 29 and August 3, \gx was observed during a pre-periastron flare (at the orbital phase approximately 0.95--1.00), as can be seen in the light curve shown in Fig. \ref{fig:lc} and that was obtained by the \textit{Swift}/BAT \citep{Gehrels04} monitor.\footnote{\url{https://swift.gsfc.nasa.gov/results/transients/}.} 
The light curves of \gx in the 2--8 energy band obtained with the \textit{IXPE} observatory as well as with \textit{SRG}/{ART-XC} in the 4--12 keV band are shown in Fig.~\ref{fig:lc1}.
The spin period for \gx was measured from the \textit{IXPE} data to be $P_{\mathrm{spin}}$=670.644(2)\,s.
The pulsed fraction in the 2--8 keV energy range of \textit{IXPE}\ was determined to be $PF = 34.0\%\pm1.1\%$, defined according to the equation $PF = (F_{\max}-F_{\min})/(F_{\max}+F_{\min})$, where $F_{\max}$ and $F_{\min}$ are the maximum and minimum count rates in the pulse profile, respectively. The pulse profiles of \gx as seen by \textit{IXPE} and {ART-XC} in different energy bands are shown in Fig.~\ref{fig:pulse-profiles-comb}. The phase-zero was chosen to coincide for the \textit{IXPE} and {ART-XC} observations, and the phase-shift was determined by comparing the pulse profiles in the 4--8 keV energy band for both observations.

\begin{figure}
\centering
\includegraphics[width=0.8\columnwidth]{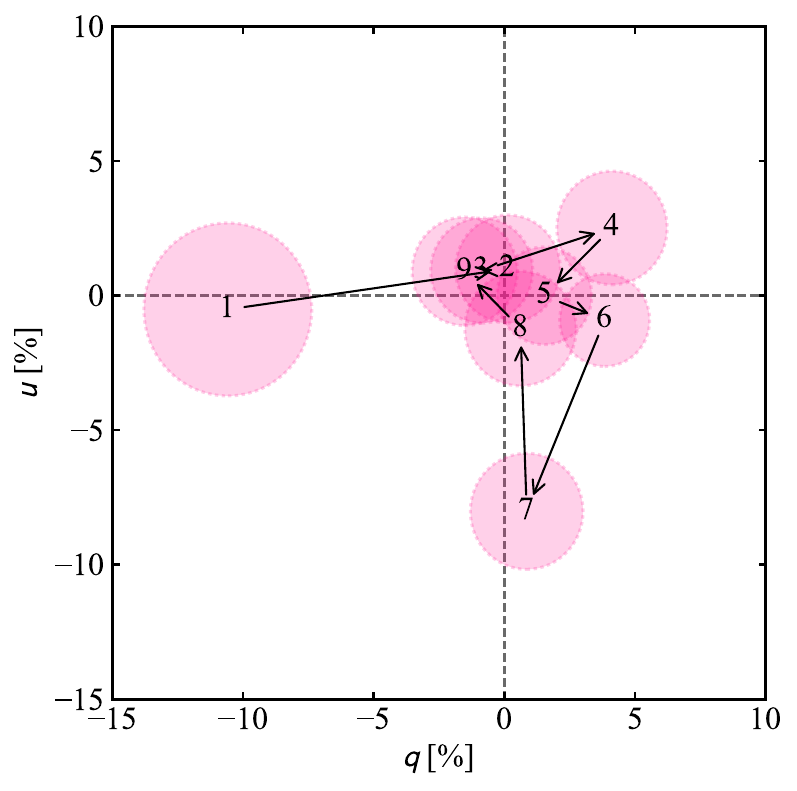}
\caption{Normalized Stokes parameters $q$ and $u$ for the phase-resolved polarimetric analysis using \texttt{pcube} (DUs combined) for the full 2--8 keV energy band. Each circle corresponds to a separate phase bin and is numbered according to its order as referenced in Table~\ref{table:phase-res-PCUBE}. The radius of the circle corresponds to a 1$\sigma$ uncertainty. }
\label{fig:Stokes_resolved}
\end{figure}

\subsection{Polarimetric analysis}

We analyzed the polarimetric properties of \gx in the model-independent way by deriving the normalized Stokes parameters $q=Q/I$ and $u=U/I$ and subsequently the PD$=\sqrt{q^2+u^2}$  and PA$=\frac{1}{2}\arctan(u/q)$ using the \texttt{pcube} algorithm (\texttt{xpbin} tool) in the \textsc{ixpeobssim} package, which follows the description by \citet{2015-Kislat} and \citet{Baldini22}.
We obtained the phase-averaged values of $q=0.7\pm0.7\%$ and $u=-0.5\pm0.7\%$ in the full 2--8 keV energy range (see Fig.~\ref{fig:Stokes_averaged} for all the DUs separately), implying that no significant polarization in the phase-averaged data is detected. 
The data were   subsequently divided into six energy bins to study a possible energy dependence of the phase-averaged polarization properties for \gx. 
The measured Stokes parameters are consistent with zero within $2\sigma$. 

\begin{figure}
\centering
\includegraphics[width=0.75\linewidth]{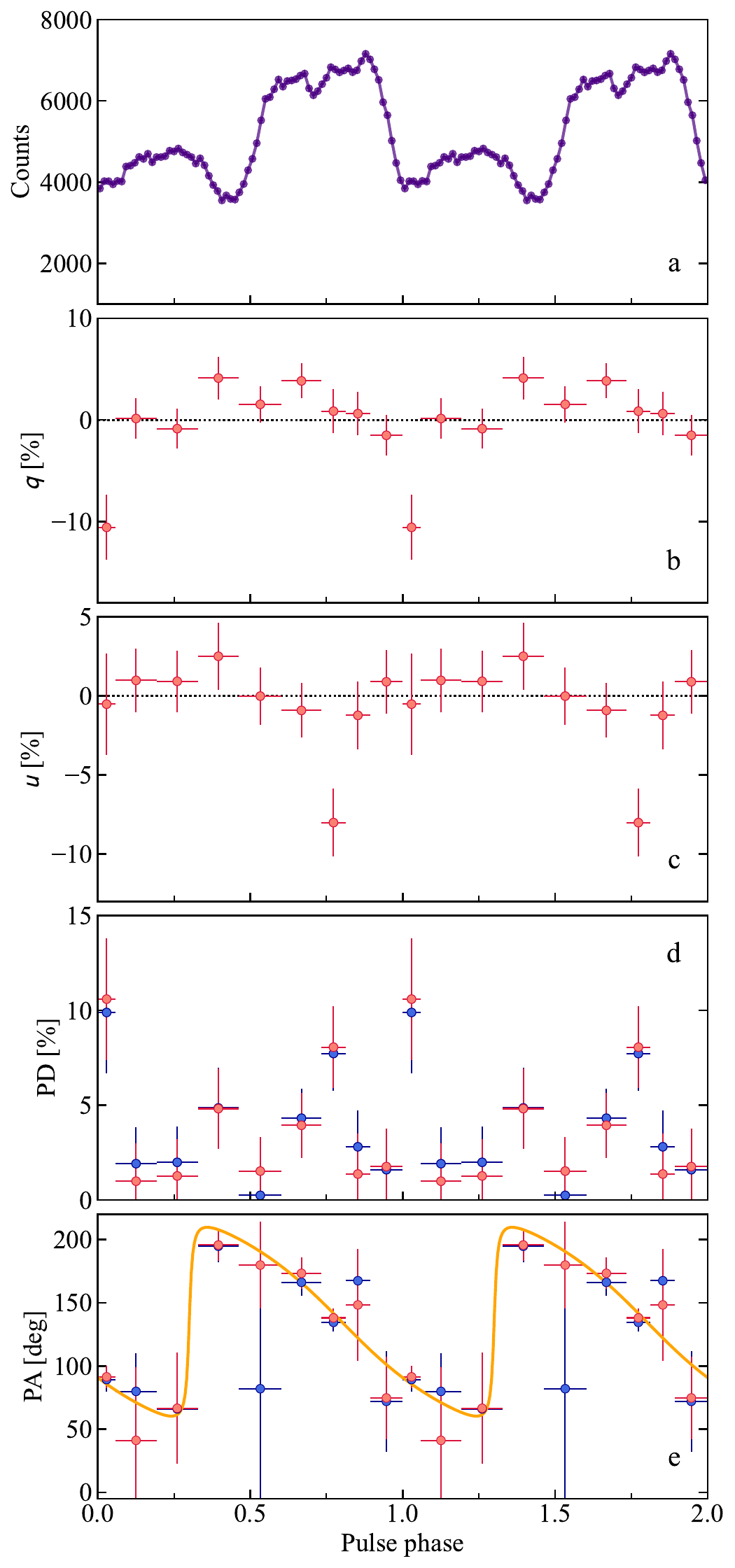}
\caption{Results from the pulse-phase-resolved analysis of \gx in the 2--8 keV range combining data from all DUs.  
\textit{Panel (a)}: Pulse profile.
\textit{Panels (b)} and \textit{(c)}: Dependence of the Stokes $q$ and $u$ parameters obtained from the \texttt{pcube} algorithm on the pulse phase. 
\textit{Panels (d)} and \textit{(e)}:  PD and PA obtained with \texttt{pcube} and from the phase-resolved spectro-polarimetric analysis using {\sc xspec} (shown by the red and blue symbols, respectively). 
The orange curve in \textit{Panel (e)} shows the best-fit RVM with $i_{\rm p}=135\degr$, $\theta=43\degr$, $\chi_{\rm p}=135\degr$, and $\phi_0/(2\pi)=-0.2$ to the PAs obtained from the pairs of  $(q,u)$ (see Sect.~\ref{sec:geom}). } 
\label{fig:phase-res-pcube-xspec}
\end{figure}

The polarization properties of XRPs are expected to be strongly variable with pulse phase, and therefore we performed a phase-resolved polarimetric analysis using the \texttt{pcube} algorithm.
The data were divided into nine separate phase bins in the full 2--8 keV range.
The results are displayed in Fig.~\ref{fig:Stokes_resolved} and in Fig.~\ref{fig:phase-res-pcube-xspec}b,c and are given in Table~\ref{table:phase-res-PCUBE}.
We found a significant (above 3$\sigma$) detection of polarization in two out of the nine phase bins and a marginal detection (above 2$\sigma$) in another two bins.

\begin{figure*}
\centering
\includegraphics[width=0.9\linewidth]{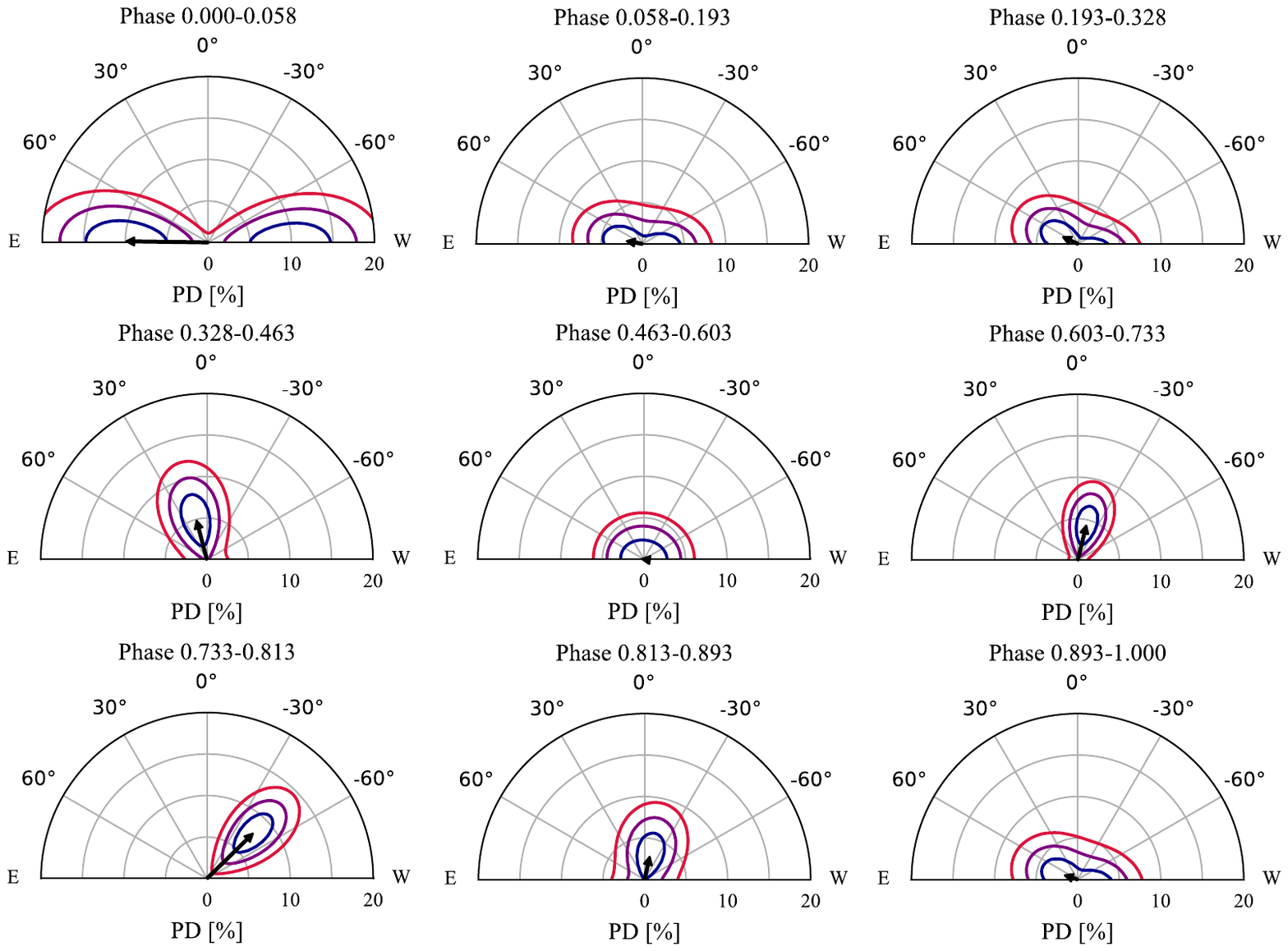}
\caption{Polarization vectors of \gx from the results of the phase-resolved spectro-polarimetric analysis. Contours at 68.3\%, 95.45\%, and 99.73\% confidence levels are shown in blue, purple, and red, respectively.}
 \label{fig:phase-res-xspec}
\end{figure*}

\begin{table}
\centering
\caption{Spectral parameters for the best-fit model obtained from the phase-averaged spectro-polarimetric analysis with \textsc{xspec}. 
}
\begin{tabular}{ccc}
\hline
\hline
 Parameter & Value & Unit\\
\hline
$N_{\mathrm{H}}$   & $24.6\pm0.4$ & $10^{22}\mathrm{\;cm^{-2}}$   \\
$E_{\mathrm{Fe}}$   & $6.35\pm 0.01$ & keV   \\
 $\sigma_{\mathrm{Fe}}$   & $0.27\pm0.03$ & keV   \\
$\mathrm{const_{DU2}}$    & $0.960\pm 0.007$ & \\
$\mathrm{const_{DU3}}$    & $0.927\pm 0.007$ & \\
hoton index   & $0.29\pm0.05$ & \\
PD & $0.67\pm0.64$ $(<2.3)$ & \% \\
PA & $-36\pm27$ & deg \\
$\mathrm{Flux_{2-8\;keV}}$ & $8.58_{-0.08}^{+0.03}$ & $\mathrm{10^{-10}\;erg\;cm^{-2}\;s^{-1}}$ \\
$\mathrm{Luminosity_{2-8\;keV}}$ & $1.3\times10^{36}$ & $\mathrm{erg\;s^{-1}}$ at $d=3.5$ kpc \\
$\chi^2$ (d.o.f.) & 1374 (1322) & \\ 
\hline
\end{tabular}
\tablefoot{
The uncertainties are given at the 68.3\% (1$\sigma$) confidence level and were obtained using \texttt{the error} command in \textsc{xspec} with $\Delta\chi^2=1$ for one parameter of interest.
The upper limit of the PD (in parentheses) corresponds to the 99\% confidence level ($\Delta\chi^2=6.635$). 
}
\label{table:best-fit}
\end{table}

\begin{table*} 
\centering
\caption{Spectro-polarimetric parameters in different pulse-phase bins obtained with \textsc{xspec}.}
\begin{tabular}{cccccc}
    \hline\hline
    Phase &  $N_{\mathrm{H}}$  & Photon index & PD & PA & $\chi^2$/d.o.f. \\ 
           & ($10^{22}\mathrm{\;cm^{-2}}$) &   & (\%) & (deg) &  \\ 
    \hline
    0.000--0.058 & $24.4\pm1.2$  & $0.15\pm0.14$ & $9.9\pm3.2$ & $89\pm9$ & 991/1017\\ 
    0.058--0.193 & $23.4\pm0.7$  & $-0.08\pm0.08$ & $1.9_{-1.9}^{+1.9}$ $(<6.8)$ & $80\pm30$ & 1262/1245\\ 
    0.193--0.328 & $25.1\pm0.7$  & $0.11\pm0.08$ & $2.0_{-1.9}^{+1.9}$ $(<6.8)$ & $66\pm34$ & 1343/1236\\ 
    0.328--0.463 & $24.3\pm0.7$  & $0.12\pm0.09$ & $4.9\pm2.1$ & $15\pm13$ & 1254/1212\\ 
    0.463--0.603 & $26.3\pm0.6$  & $0.45\pm0.07$ & $0.2_{-0.2}^{+1.7}$ $(<4.6)$ & $82\pm90$ & 1321/1269\\ 
    0.603--0.733 & $25.6\pm0.6$  & $0.30\pm 0.06$ & $4.3\pm1.6$ & $-14\pm11$ & 1398/1278\\ 
    0.733--0.813 & $24.8\pm0.7$  & $0.24\pm0.08$ & $7.7\pm2.0$ & $-46\pm7$ & 1271/1206\\ 
    0.813--0.893 & $26.5\pm0.7$  & $0.59\pm0.08$ & $2.8\pm1.9$ & $-13\pm22$ & 1273/1212\\ 
    0.893--1.000 & $26.8\pm0.7$  & $0.59\pm0.08$ & $1.6_{-1.6}^{+1.9}$ $(<6.5)$ & $72\pm40$ & 1241/1227\\  
    \hline
    \end{tabular}
\tablefoot{The uncertainties computed using the \texttt{error} command 
are given at the 68.3\% (1$\sigma$) confidence level ($\Delta\chi^2=1$ for one parameter of interest). 
The upper limits of the PD (in parentheses) correspond to the 99\% confidence level ($\Delta\chi^2=6.635$) and are quoted when the PD is consistent with zero within 1$\sigma$.} 
\label{table:phase-res}
\end{table*}

Next, to fully account for the energy dispersion and spectral shape, we performed a model-dependent spectro-polarimetric analysis by simultaneously fitting the Stokes $I$, $Q$, and $U$ spectra in {\sc xspec} in order to study the polarization properties as a function of energy.
The continuum spectra of XRPs are typically described by purely phenomenological spectral models, such as a power law with a cutoff at high energies. 
Several models of varying complexity have been used to describe the spectral continuum of \gx.
However, for the energy resolution and energy range of \textit{IXPE}, we used a more simplified model to fit the spectra.
This spectral model consists of an absorbed power law (for the absorption we used \texttt{tbabs} model with the abundances from \citealt{Wilms2000}). 
Additionally, a prominent iron emission line (complex) was also observed, and therefore a Gaussian component was introduced.
A re-normalization constant was used to account for possible discrepancies between the separate DUs (\texttt{const}), and for DU1 it was fixed at unity.
Finally, the \texttt{polconst} polarization model, assuming a constant PD and PA with energy, was applied to the power law, while the Gaussian component was assumed to be unpolarized. 
The resulting model,
\begin{eqnarray*} 
\texttt{ tbabs$\times$(polconst$\times$powerlaw+gaussian)$\times$const},
\label{eq:spec-model}
\end{eqnarray*}
was used to fit both the phase-averaged and the phase-resolved data. A systematic error of 5\% was added over the entire 2--8 keV energy range for the phase-averaged spectro-polarimetric analysis.
The results of the spectral fitting of the phase-averaged data are given in Table~\ref{table:best-fit}.
For the phase-averaged data, we did not detect any significant polarization, with a 99\% upper limit of PD$<2.3$\%.
The \texttt{polconst} polarization model was replaced with the \texttt{pollin} (linear energy dependence of the PD and PA) and \texttt{polpow} (power-law energy dependence of the PD and PA) models; however, neither one gave an improvement of the fit. Thus, we found no evidence for an energy-dependent polarization.

The same model used for the phase-averaged spectro-polarimetric analysis was used to fit the phase-resolved spectra. However, the cross-calibration constants for DU2 and DU3 were fixed to the best-fit values obtained for the phase-averaged analysis (see Table~\ref{table:best-fit}).
Due to low statistics in the phase-resolved spectra, the energy and width of the Gaussian component were fixed at the best-fit values from Table~\ref{table:best-fit}. 
The \texttt{steppar} command in {\sc xspec} was used to obtain the constraints  on polarization parameters. 
The resulting PD and PA are shown in  Fig.~\ref{fig:phase-res-pcube-xspec}d,e and given in Table~\ref{table:phase-res}. 
The two different approaches using \texttt{pcube} and \textsc{xspec} gave compatible results.
The 2D contour plots at 68.3\%, 95.45\%, and 99.73\% confidence levels for the  PD and PA pair are presented in  Fig.~\ref{fig:phase-res-xspec}.

\begin{figure*}
\centering
\includegraphics[width=0.5\linewidth]{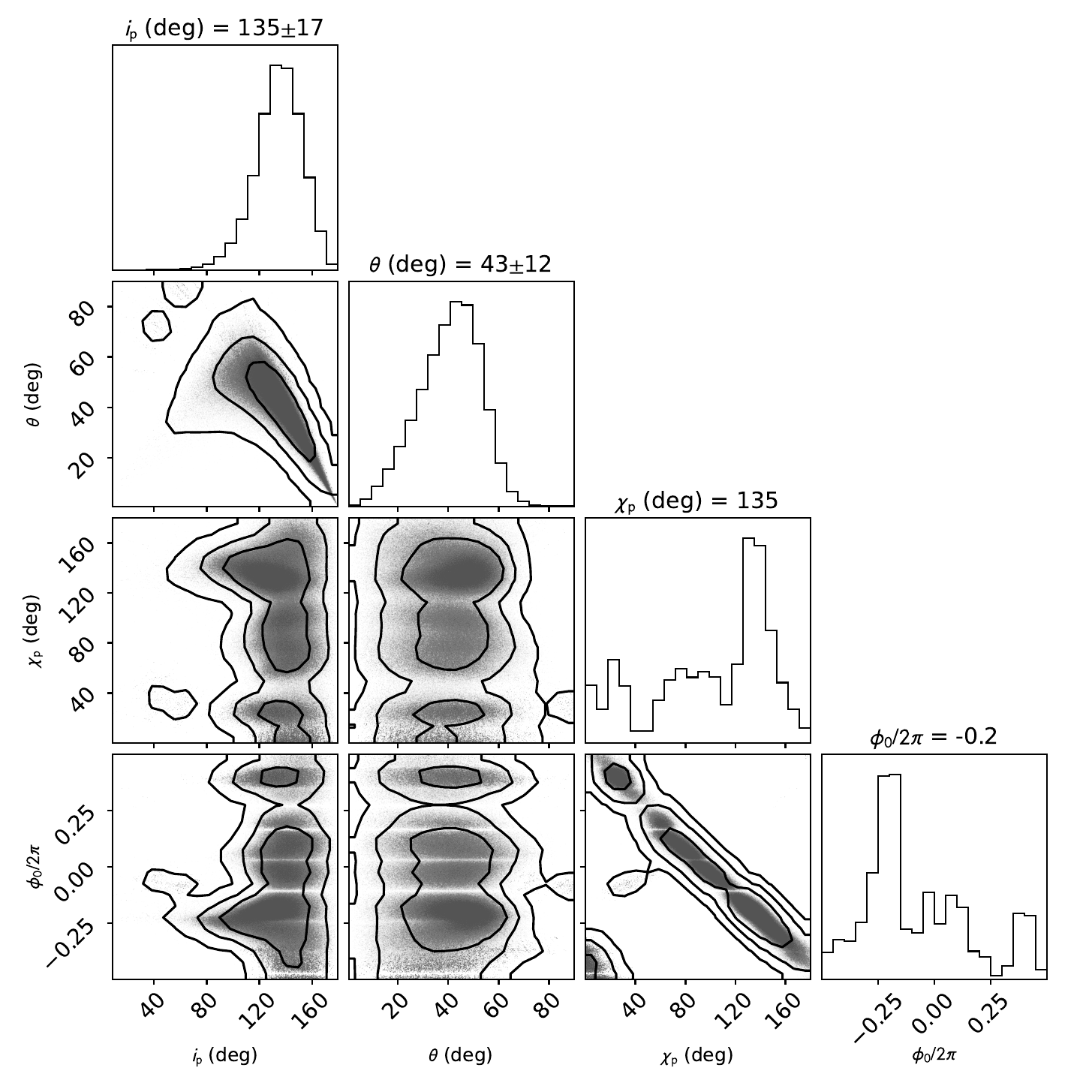}
\caption{Corner plot of the posterior distribution for parameters of the RVM  model fitted directly to the $(q,u)$ values using the likelihood function \eqref{eq:PA_dist}. 
The two-dimensional contours correspond to 68.3\%, 95.45\%, and 99.73\%  confidence levels. 
The histograms show the normalized one-dimensional distributions for a given parameter derived from the posterior samples. 
}
\label{fig:corner_SST}
\end{figure*}
\begin{figure*}
\centering
\includegraphics[width=0.7\linewidth]{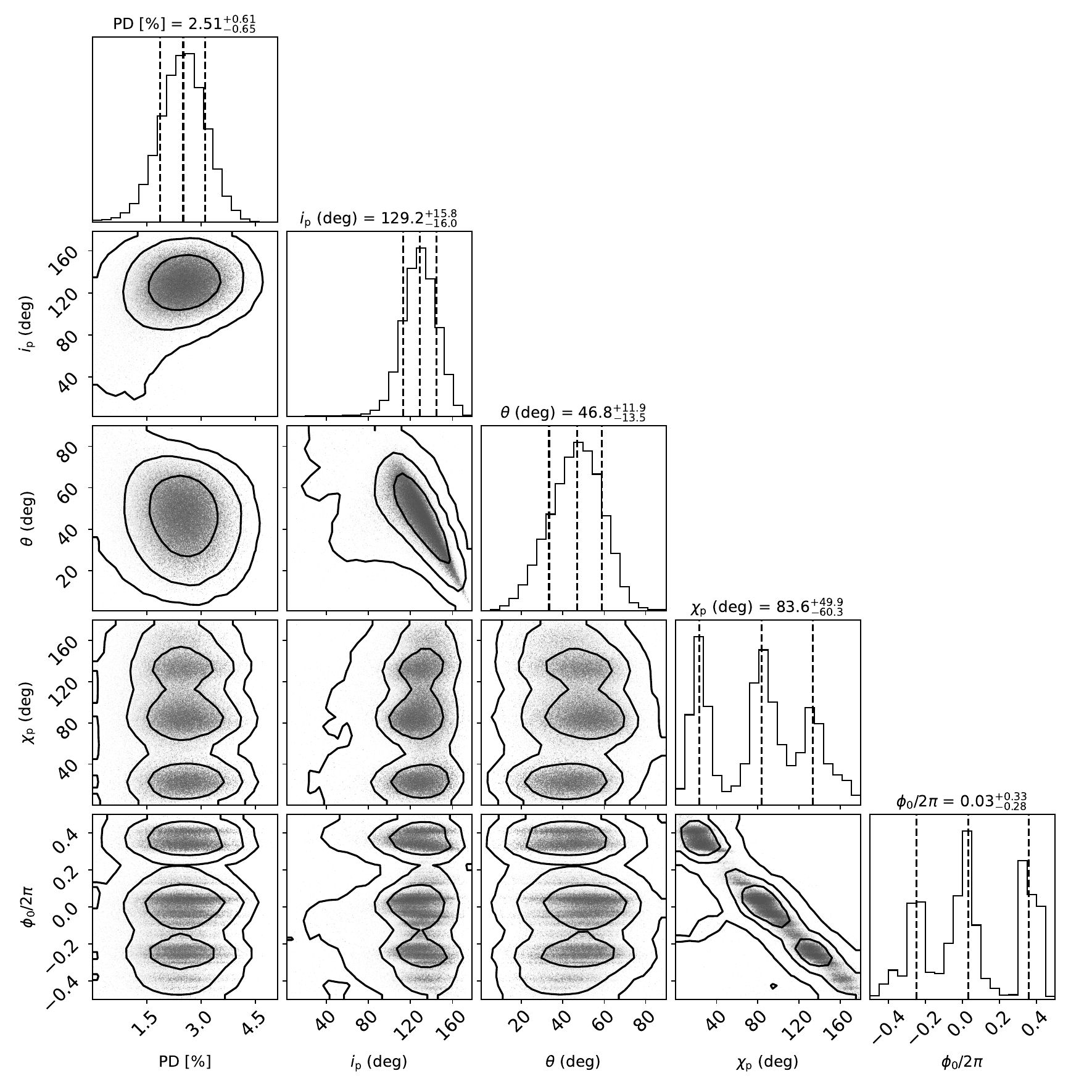}
\caption{Same as Fig.~\ref{fig:corner_SST} but obtained via unbinned analysis \citep{Gonzalez2023}. }
 \label{fig:corner_JH}
\end{figure*}

\section {Discussion}
\label{sec:discussion} 

\subsection{Pulsar geometry} 
\label{sec:geom}

The rotating vector model (RVM; \citealt{Radhakrishnan69})  can be used to constrain the pulsar geometry. 
Assuming that radiation is dominated by the ordinary mode photons (O-mode), the PA is given by equation\,(30) in \citet{Poutanen2020}: 
\begin{equation} \label{eq:pa_rvm}
\tan (\mbox{PA}\!-\!\chi_{\rm p})\!=\! \frac{-\sin \theta\ \sin (\phi-\phi_0)}
{\sin i_{\rm p} \cos \theta\!  - \! \cos i_{\rm p} \sin \theta  \cos (\phi\!-\!\phi_0) } ,
\end{equation} 
where $\chi_{\rm p}$ is the position angle (measured from north to east) of the pulsar angular momentum, $i_{\rm p}$ is the inclination of the pulsar spin to the line of sight, $\theta$ is the angle between the magnetic dipole and the spin axis, and $\phi_0$ is the phase when the northern magnetic pole passes in front of the observer.   
If radiation escapes predominantly in the extraordinary mode (X-mode), the position angle of the pulsar angular momentum is $\chi_{\rm p}\pm90\degr$.
In the non-relativistic RVM, the PA does not depend on the PD of the radiation escaping from the NS surface, or any other details. In fact, the polarization plane rotates when radiation travels through  the NS magnetosphere up to the adiabatic radius, a few tens of the NS radius \citep{2003MNRAS.342..134H}. At such distances, the dipole field component dominates, and under these circumstances, the RVM model becomes applicable. 
The general relativistic effects influence the polarization plane only if the NS rotates rapidly  \citep[see][]{Poutanen2020}. 
 
In spite of the fact that the observed PAs are not well determined in many phase bins, we can use measurements of the Stokes $q$ and $u$ parameters (which are normally distributed) in all phase bins to get  constraints on RVM parameters. 
For any $(q,u)$ and their error $\sigma_{\rm p}$, the probability density function of the PA, $\psi$, can be computed as \citep{Naghizadeh1993}: 
\begin{equation} \label{eq:PA_dist}
G(\psi) = \frac{1}{\sqrt{\pi}} 
\left\{  \frac{1}{\sqrt{\pi}}  + 
\eta {\rm e}^{\eta^2} 
\left[ 1 + {\rm erf}(\eta) \right]
\right\} {\rm e}^{-p_0^2/2}.
\end{equation}
Here, $p_0=\sqrt{q^2+u^2}/\sigma_{\rm p}$ is the ``measured'' PD in units of the error,  $\eta=p_0 \cos[2(\psi-\psi_0)]/\sqrt{2}$, $\psi_0=\frac{1}{2}\arctan(u/q)$ is the central PA obtained from the Stokes parameters, and \mbox{erf} is the error function.

We fit the RVM to the pulse-phase dependence of the $(q,u)$ obtained from \texttt{pcube} using the affine invariant Markov chain Monte Carlo (MCMC) ensemble sampler {\sc emcee} package of {\sc python} \citep{emcee13} and applying the likelihood function $L= \Pi_i  G(\psi_i)$ with the product taken over all phase bins.  
The covariance plot for the RVM parameters is shown in Fig.~\ref{fig:corner_SST}.   
We note  that fitting the PA values obtained from \textsc{xspec} with the RVM and assuming Gaussian errors and $\chi^2$ statistics gives nearly identical results. 

It is clear that the inclination and the magnetic obliquity are rather well determined:  $i_{\rm p}=135\degr\pm17\degr$ and $\theta=43\degr\pm12\degr$. 
However, other parameters allowed for multiple solutions. 
The reason for this is simple: 
Most of the detections  of polarization are only marginal.
The posterior distribution of $\chi_{\rm p}$ has three broad peaks at 25\degr, 80\degr, and the strongest one at 135\degr. 
These peaks correspond to the peaks in $\phi_0/(2\pi)$ at 0.4, 0.0, and $-0.2$, respectively.

\begin{figure}
\centering
\includegraphics[width=0.7\linewidth]{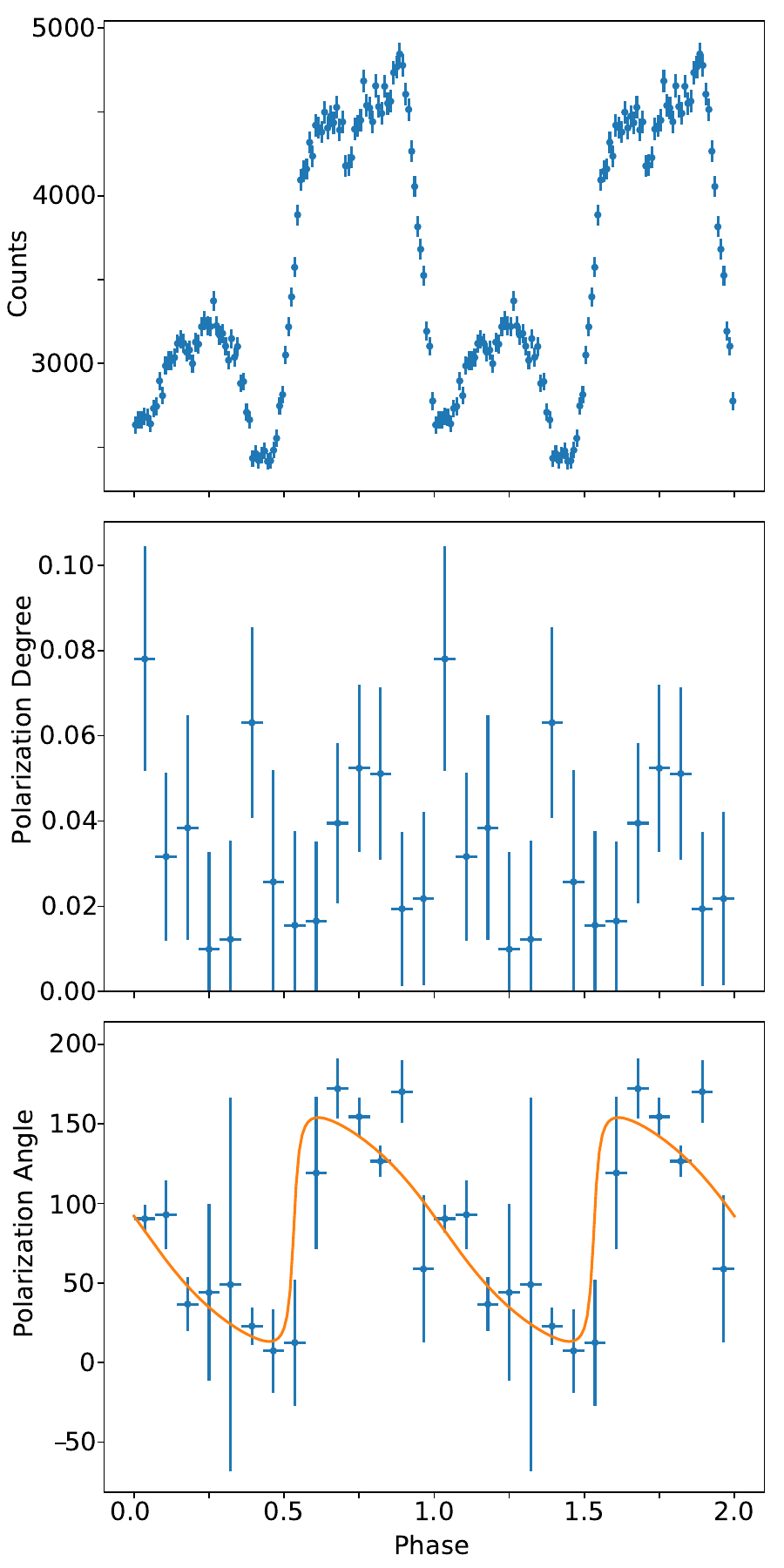}
\caption{Pulse profile, PD, and PA obtained from the maximum-likelihood analysis \citep{Gonzalez2023}.  
The error bars give the most likely values of PD and PA with $\Delta \ln L=0.5$ (1$\sigma$) confidence regions. 
The orange curve at the bottom panel shows the best-fit RVM solution with the following parameters $i_{\rm p}=129\degr$,   $\theta=47\degr$, $\chi_{\rm p}=84\degr$, $\phi_0/(2\pi)=0.03$.  
}
 \label{fig:unbinned}
\end{figure}

As a test of our RVM fit to the phase-binned data, we applied an alternative unbinned photon-by-photon analysis \citep{Gonzalez2023}.
We ran MCMC simulations to get the estimates on the RVM parameters.
The results are shown in Fig.~\ref{fig:corner_JH}.  
The photoelectron emission angles for each event were used to constrain an RVM that predicts the distribution of photoelectron angles as a function of the phase.  
The PA of the best-fit RVM is depicted by an orange curve in Fig.~\ref{fig:unbinned}. 
Furthermore, for illustration, we obtained a maximum-likelihood estimate of the PD and PA, assuming constant polarization in each of the fourteen phase bins, as shown by the error bars in the lower two panels of Fig.~\ref{fig:unbinned}.  
The mean PD in the frame of the magnetic pole is about 2.5\%.
The best-fit $i_{\rm p}=129\degr\pm16\degr$ and  $\theta=47\degr\pm 12\degr$ are in perfect agreement with the RVM fit to the phase-binned data. 
However, the position angle of the pulsar spin and the zero phase $\phi_0$  are not well determined. 
Here too, the posterior distributions show multiple peaks and strong correlation between these parameters.  
Three peaks in the distributions of $\chi_{\rm p}$ and $\phi_0$ at positions close to those of the phase-binned RVM fits can be seen.

\subsection{Scattering by the stellar wind}
\label{sec:wind}

The relatively low PD detected for \gx is in line with other XRPs observed with \textit{IXPE} (Cen X-3, Her X-1, GRO J1008$-$57, Vela X-1, 4U 1626$-$67, EXO 2030+375). 
The low PD is in contradiction with previous theoretical predictions, which estimate PDs as high as 80\% \citep{1988ApJ...324.1056M,CH1_21}.
We find it interesting that more luminous XRPs  ($L_{\rm X} \ge 10^{37}$\,erg\,s$^{-1}$) demonstrate a significantly higher phase-averaged PD, up to 6\%--10\% in Her X-1 \citep{Doroshenko22} and 4\%--6\% in Cen X-3 \citep{Tsygankov22}, than the less luminous XRPs. 
It can be connected with the difference in the physical properties of the radiation regions. 
In particular, \citet{Doroshenko22} proposed a model to explain the low PD by means of an overheated upper NS atmosphere with a sharp temperature gradient between the overheated upper atmospheric layer and the underlying cooler layer.
The most important component of this model is the point of the so-called vacuum resonance, where the mode conversion occurs.
If the point of vacuum resonance is located at the transition between the hot and cool layer, this may result in a low PD. 
The described model can be correct for relatively low local accretion rates only.

In addition to an intrinsic low polarization, we considered the possibility that the X-ray radiation can be depolarized in the plasma surrounding the XRP because \gx is significantly more obscured than other XRPs. 
A detailed qualitative analysis of the reasons for depolarization was presented in \citet{Tsygankov22}.
Here, we considered the depolarization of the X-ray flux passing  through the surrounding plasma in more detail.
The XRP \gx is subject to the dense stellar wind of its massive companion star. This wind is characterized by a relatively slow velocity and a huge mass-loss rate. 
Scattering in the wind is expected to reduce intrinsic polarization.
Whether or not the wind is able to affect the observed polarization depends on the wind Thomson optical depth, which can be estimated as \citep{2015-Kallman} 
$\tau_{\rm T} \simeq 2\times10^{-4}\dot{M}_{8}a_{12}^{-1}v_{\rm x8}^{-1}$, where $\dot{M}_8$ is the  mass-loss rate in units $10^{-8}M_{\odot}\,\mathrm{yr}^{-1}$, 
$a_{12}$ is the separation in units $10^{12}$\,cm, and $v_{\rm x8}$ is the  wind velocity at the X-ray source in units $1000\,\mathrm{km\,s}^{-1}$.  
For \gx, taking typical values for the wind parameters $\dot{M}_8\sim10^{3}$ and $v_{\rm x8}=0.3$ as well as the orbital separation at periastron passage $a=a_{\rm p}\sim 95R_{\odot}$ (i.e., $a_{12}=6.7$), we obtained the optical depth through the wind of $\tau_{\rm T}\sim0.1$.

Our estimate was made under the assumption of a homogeneous, spherically symmetric wind. 
However, the orbiting  NS disturbs the stellar wind, leading to various  inhomogeneities in the vicinity of the XRP. 
As a result, the optical depth along the line of sight can be larger. 
In the first approximation, this optical depth can be estimated from $N_{\rm H},$ assuming that the number of electrons along the line of sight is the same.  
For the $N_{\rm H}$ in the range $(5-20)\times 10^{23}$\,cm$^{-2}$ 
\citep{Furst18,Armin19, Doroshenko.etal.2010, Ji21}, we obtained $\tau_{\rm T} \approx \sigma_{\rm T}\, N_{\rm H} \approx $0.3--1.2.

 \begin{figure}
\centering
\includegraphics[width=0.99\linewidth]{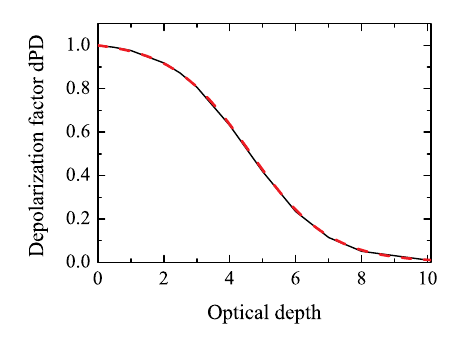}
 \caption{Dependence of the depolarization factor on the plasma optical depth. The fit with the expression given by Eq.~\eqref{depol} is also shown with the red dashed curve. 
 }
  \label{fig:depol}
 \end{figure}

In this paragraph, we describe our consideration of how the polarized radiation will be depolarized when it is transmitted through the ionized plasma slab. 
We neglected emission by the plasma itself and assumed that the slab is situated far enough from the X-ray source so that the external flux is directed normally to the slab surface. 
We then computed polarization properties of radiation that is transmitted through the slab along its normal.
We assumed that the incoming radiation is 100\% linearly polarized. 
We solved the radiation transfer equation accounting for polarization under electron scattering  \citep{chandrasekhar1960,Sobolev63,Suleimanov23}. 
We defined the depolarization factor dPD=PD$'$/PD as the ratio of the PD of the transmitted radiation to that of the incident one. 
The results of our calculations for different optical depths of the slab are shown in Fig.~\ref{fig:depol}.
It is clear that if the slab is almost transparent ($\tau <1$), only a small part of the radiation is scattered, and the PD decreases insignificantly. 
At higher optical depths, the fraction of scattered radiation increases, resulting in a smaller PD, and finally the PD becomes consistent with the case of the pure electron scattering atmosphere \citep{chandrasekhar1960,Sobolev63} and approaches zero because of the axial symmetry. 
The dependence of dPD on the slab optical depth $\tau$ is well approximated by the expression  
\be \label{depol}
    {\rm dPD} = A\ \left[\exp\left(\frac{\tau-\tau_*}{\Delta \tau}\right)+1\right]^{-1},
\ee
where $\tau_*=4.6$ is the optical depth providing dPD=0.5, $\Delta \tau = 1.2$ is a smearing parameter, and 
$A=\exp\left(-\tau_*/\Delta \tau\right)+1\approx 1.02$ is a normalization constant.
We note that $\tau$ is the total optical depth including not only electron scattering but also the free-free and bound-free absorption. 
Figure~\ref{fig:depol} shows that, in principle, the PD can decrease by a factor of two for $\tau\sim 5$, but such a large optical depth of the envelope  around the XRP is clearly too large and would lead to a dramatic decrease of the pulsed fraction, which is not observed. 
Thus, we can conclude that scattering in the wind is not a likely source of a relatively low PD. 

In addition to scattering, the wind may also reprocess some radiation of the XRP. 
In particular, the strength of fluorescent iron lines K$\alpha$ and K$\beta$ observed from \gx \citep[see, e.g.,][]{Ji21} correlates with the X-ray flux and with the local absorption. 
The lines contribute about 10\% of the flux in the \textit{IXPE} band.  
Thus, these unpolarized lines can produce a depolarization factor of about 0.9, which means that they cannot significantly depolarize the XRP radiation.

\section{Summary}
\label{sec:summary} 

The XRP \gx was observed by \textit{IXPE} between 2022 July 29 and August 3 during a pre-periastron flare. Simultaneous observations of \gx were carried out by the {ART-XC} telescope between 2022 July 30 and August 1.
The main results of our study of the polarimetric properties of \gx are summarized as follows:
\begin{enumerate}
\item We have not detected a significant polarization  in the phase-averaged data with an upper limit to the PD of 2.3\% (at 99\% confidence level) in the full 2--8 keV energy range of \textit{IXPE}. Likewise, no significant polarization was detected in any energy bin in the phase-averaged data. 
\item The phase-resolved polarimetric analysis revealed a significant detection of polarization in two out of nine phase bins and a marginal detection in three bins, with the PD ranging from $\sim$3\% to $\sim$10\% and the PA varying in a very wide range from $\sim$0\degr\ to $\sim$160\degr.
\item  Application of the RVM to the phase-resolved PA (accounting even for the points with low significance) allowed us to estimate the inclination of the NS rotation axis to the line of sight of $i_{\rm p} \approx 130\degr-140\degr$ and the magnetic obliquity of $\theta \approx 40\degr-50\degr$. 
The measured pulsar spin inclination is in good agreement with the orbital inclination of 52\degr--78\degr, with some preference toward lower values determined by \citet{2008-Leahy-Kostka}. 
We note that the preferred lowest orbital inclination is equivalent to 128\degr. 
Thus, it is possible that the rotation axis of the pulsar is nearly aligned with the orbital axis. 
On the other hand, the NS in the system might rotate retrogradely, as was proposed by \citet{juhani20}. 
The position angle of the pulsar spin and the zero phase, when the northern magnetic pole passes in front of the observer, allow for multiple solutions. 
We find excellent agreement between phase-binned and unbinned analyses. 
  
\item The contribution of the reprocessed radiation and scattering in the surrounding matter can depolarize the intrinsic XRP flux by no more than 10\% to 15\%.
\end{enumerate}
Additional observations of \gx with \textit{IXPE} may improve the statistics, allowing us to get better constraints on the pulsar axis position angle and the zero phase $\phi_0$.

\begin{acknowledgements}
The Imaging X-ray Polarimetry Explorer is a joint US and Italian mission.  
The US contribution is supported by the National Aeronautics and Space Administration (NASA) and led and managed by its Marshall Space Flight Center (MSFC), with industry partner Ball Aerospace (contract NNM15AA18C).  
The Italian contribution is supported by the Italian Space Agency (Agenzia Spaziale Italiana, ASI) through contract ASI-OHBI-2017-12-I.0, agreements ASI-INAF-2017-12-H0 and ASI-INFN-2017.13-H0, and its Space Science Data Center (SSDC) with agreements ASI-INAF-2022-14-HH.0 and ASI-INFN 2021-43-HH.0, and by the Istituto Nazionale di Astrofisica (INAF) and the Istituto Nazionale di Fisica Nucleare (INFN) in Italy.
This research used data products provided by the IXPE Team (MSFC, SSDC, INAF, and INFN) and distributed with additional software tools by the High-Energy Astrophysics Science Archive Research Center (HEASARC), at NASA Goddard Space Flight Center (GSFC). 

This research has been supported by Deutsche  Forschungsgemeinschaft (DFG) grant WE 1312/59-1 (VFS). 
We also acknowledge support from the German Academic Exchange Service (DAAD) travel grant 57525212 (VFS, VD),  
the Academy of Finland grants 333112, 349144, 349373, and 349906 (SST, JP),  
the V\"ais\"al\"a Foundation (SST), 
CNES fellowship grant (DG-C), 
Natural Sciences and Engineering Research Council of Canada (JH), 
the Russian Science Foundation grant 19-12-00423 (AAL, SVM, AES), 
and UKRI Stephen Hawking fellowship (AAM).  
\end{acknowledgements}

\bibliographystyle{yahapj}
\bibliography{46994ref}

\end{document}